\documentclass[11pt]{article}
\usepackage{latexsym,amssymb,amsmath}
\usepackage[dvips]{graphics,graphicx,epsfig,color}
\begin{document}
\begin{center}
 {\bf\Large Synthesis of unilateral radiators}
\end{center}
\begin{center}
{Armand WIRGIN$^{1}$ \\~\\ {\it $^{1}$ Laboratoire de M\'ecanique
et d'Acoustique, UPR 7051 du CNRS, Marseille, France.}}
\end{center}
~\\\\\\

{\small{\bf Abstract.}~- A radiator is typically a parabolic
mirror illuminated by an electromagnetic source, or a cylindrical
transducer of resonant vibrations. Both of these devices are
designed to radiate either a beam of parallel rays or a (focused)
beam that converges to a point or a line. Consequently, at the
worst, the radiation pattern is largely restricted to a {\it half
space}, and at the best, to a cone or cylinder-like subspace of
this half space. Such devices can therefore be termed {\it
unilateral radiators}. This study is devoted to the synthesis of
the sources that can give rise to such radiation, the underlying
motivation being  the removal of the material presence of the
mirror or transducer casing from which waves coming from other
boundaries could reflect or diffract.}
%%%%%%%%%%%%%%%%%%%%%%%%%%%%%%%%%%%%%%%%%%%%%%%%%%%%%%%%%%%%%%%%%%%%%%%
\newpage
%%%%%%%%%%%%%%%%%%%%%%%%%%%%%%%%%%%%%%%%%%%%%%%%%%%%%%%%%%%%%%%%%%%%%%%%
\tableofcontents
%%%%%%%%%%%%%%%%%%%%%%%%%%%%%%%%%%%%%%%%%%%%%%%%%%%%%%%%%%%%%%%%%%%%%%%%
\newpage
%%%%%%%%%%%%%%%%%%%%%%%%%%%%%%%%%%%%%%%%%%%%%%%%%%%%%%%%%%%%%%%%%%%%%%%%
\section{Introduction}\label{intro}
Often one wants to predict the wavefield arising from sources
radiating waves that are diffracted from various material objects
(termed {\it obstacles}) in an otherwise homogeneous space. In
real life, the so-called radiated wave is the result of a complex
process involving conversion of an electrical signal into a (e.g.,
acoustic) wave which is formed in some manner within a so-called
antenna (which sometimes reduces to a casing or radome). The
radiated wave then propagates towards the obstacles present in the
space and is diffracted by the latter. Some of the diffracted
waves are redirected towards the antenna, and since the latter is
also a material object, it can diffract these waves in its turn.

It is difficult to account for this multiple diffraction between
the antenna and the objects, so the usual procedure is to assume
that the radiator is only a collection of sources without material
presence. In other words, one assumes that the antenna radiates
waves, but does not diffract waves returning from the obstacles.
\begin{figure}
[ptb]
\begin{center}
\includegraphics[scale=0.6] {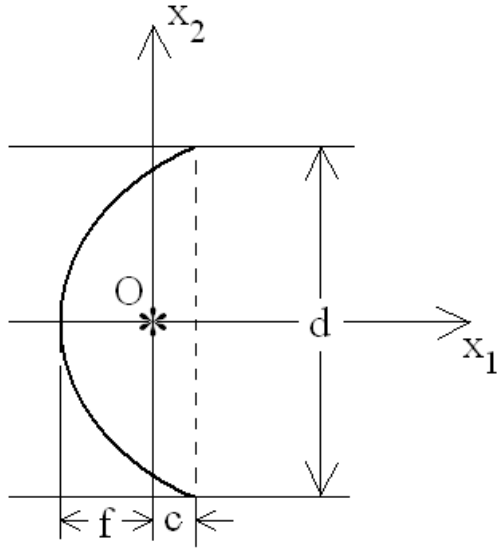}
  \caption{Cross section view of a parabolic cylinder  radiator fed by a line
  source situated at the focal point $O$.}
  \label{fig0}
  \end{center}
\end{figure}

In order for this to be possible, at least in theory, one must
reduce the radiator, which is composed of sources and an antenna
(see for instance, fig. \ref{fig0} for an example of a parabolic
cylinder radiator), to a mere collection of sources. The latter,
by definition, radiate outgoing waves, but do not diffract
incoming waves. Moreover, in the present instance, one wants the
radiator to radiate unilaterally, i.e., predominantly within a
half space. Usually, this is not a simple task, as a point source
radiates in all space, so that it is natural to think that a
collection of point sources will also radiate in all space.

In the present investigation, we  show that a particular
combination of so-called single and double sources enables one to
synthesize an essentially-unilateral radiator.
%%%%%%%%%%%%%%%%%%%%%%%%%%%%%%%%%%%%%%%%%%%%%%%%%%%%%%%%%%%%%%%%
\section{Diffraction of a wave by an impenetrable screen with a
finite-size aperture}\label{screen}
%
%%%%%%%%%%%%%%%%%%%%
\subsection{Preliminaries}
This section is devoted to the study of the simplest of radiating
devices: an impenetrable screen with an aperture.

Henceforth, we treat only 2D problems in which the support of the
sources are cylinders parallel to the $x_{3}$ cartesian coordinate
and the objects do not depend on this coordinate either.

\begin{figure}
[ptb]
\begin{center}
\includegraphics[scale=0.4] {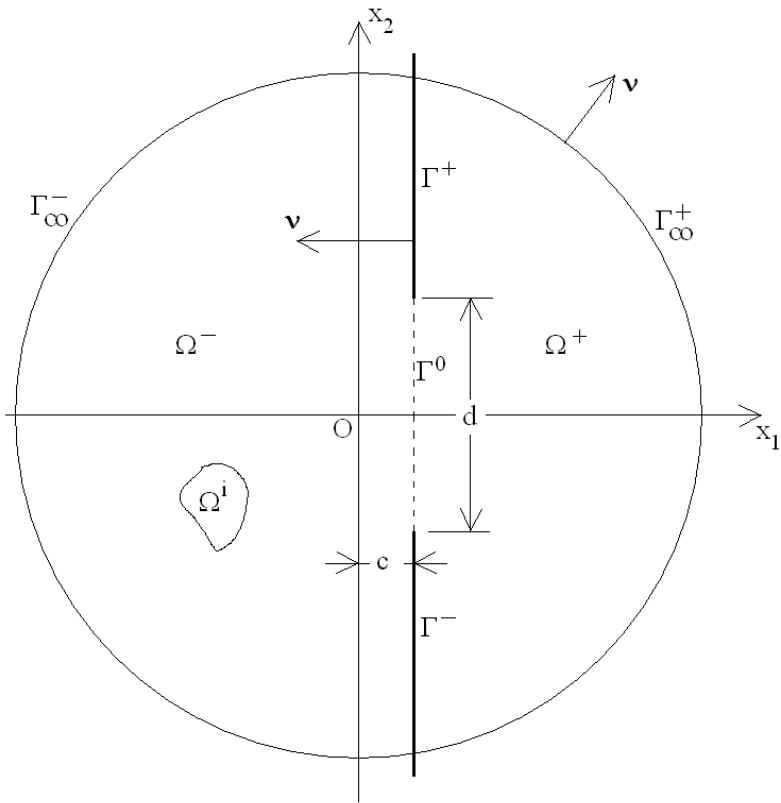}
  \caption{Cross section view of the configuration in which the wave radiated
  by a cylindrical source is diffracted by a slit aperture in an impenetrable screen.}
  \label{fig1}
  \end{center}
\end{figure}

The configuration, in the $x_{1}-x_{2}$ plane, is depicted in fig.
\ref{fig1}. The infinitely-thin impenetrable screen is denoted by
the vertical dark black lines $\Gamma^{+}$ and $\Gamma^{-}$ which
are separated by a slit-like aperture $\Gamma^{0}$. This obstacle
is illuminated by a wave radiated by sources contained within the
domain $\Omega^{i}$. The half space to the left of the screen
(minus the support of the sources) is designated by $\Omega^{-}$
and the half space to the right of the screen by $\Omega^{+}$.
These half spaces can be thought of as being closed by the semi
circles (of infinite radius) $\Gamma_{\infty}^{-}$
$\Gamma_{\infty}^{+}$ respectively. The outward-pointing unit
vector normal to $\Gamma^{-}$, $\Gamma^{0}$,
$\Gamma_{\infty}^{-}$, $\Gamma_{\infty}^{+}$ is designated by
$\boldsymbol{\nu}$, as in fig. \ref{fig1}.

The horizontal distance of the screen from the origin $O$ is $c$
and the width of the slit is $d$, with the $x_{1}$ axis being at
equal vertical distance from the two extremities of the slit.

Henceforth, we shall be concerned with a scalar wave problem such
as one that arises in acoustics (in fluids). The total scalar
(pressure) wavefield will be represented by the function
$u(\mathbf{x},\omega)$ in the space-frequency domain, with
$\mathbf{x}=(x_{1},x_{2})$ the position vector in the
cross-section plane and $\omega$ the angular frequency. The
implicit temporal factor is $\exp(-i\omega t)$, with $t$ the time
variable.  The  wavefield does not depend on $x_{3}$ due to the
fact that the sources and the screen+slit do not depend on this
variable.
%%%%%%%%%%%%%%%%%%%%%%%%
\subsection{The field in the absence of the screen}
In the absence of the screen, the total field is only the one
radiated by the applied sources so that the problem reduces to
determining $u(\mathbf{x},\omega)$ which satisfies
\begin{equation}\label{screen1}
(\triangle+k^{2})u(\mathbf{x},\omega)=-s(\mathbf{x},\omega)~~,~~\mathbf{x}\in\mathbb{R}^{2}~,
\end{equation}
(wherein $s(\mathbf{x},\omega)$ is the source density,
$k=\frac{\omega}{v}$  the wavenumber, and $v$ the velocity in the
fluid medium, assumed at present to occupy all of
$\mathbb{R}^{2}$),
\begin{equation}\label{screen2}
u(\mathbf{x},\omega)\sim \text{outgoing
waves}~;~\|\mathbf{x}\|\rightarrow\infty~~,~~\mathbf{x}\in\mathbb{R}^{2}~.
\end{equation}
If $s(\mathbf{x},\omega)=\delta(\mathbf{x}-\mathbf{x}')$, wherein
$\mathbf{x}'=(x'_{1},x'_{2})$ defines the position of the line
source and $\delta(~)$ is the Dirac delta distribution, then
\begin{equation}\label{screen3}
u(\mathbf{x},\omega):=G(\mathbf{x},\mathbf{x}',\omega)=\frac{i}{4}
H_{0}^{(1)}(k\|\mathbf{x}-\mathbf{x}'\|)
~~,~~\mathbf{x}\in\mathbb{R}^{2}~,
\end{equation}
with $H_{0}^{(1)}(~)$  the zeroth order Hankel function of the
first kind.

For other source distributions, one finds
\begin{equation}\label{screen4}
u(\mathbf{x},\omega)=\int_{\mathbb{R}^{2}}G(\mathbf{x},\mathbf{x}',\omega)
s(\mathbf{x}',\omega)d\varpi(\mathbf{x}')
~~;~~\mathbf{x}\in\mathbb{R}^{2}~,
\end{equation}
wherein $d\varpi(\mathbf{x}')$ is the infinitesimal element of
area in the $x_{1}-x_{2}$ plane.
%%%%%%%%%%%%%%%%%%%%%%%%
\subsection{The field in the presence of the screen}
In the presence of the screen, the field of (\ref{screen4}) no
longer constitutes the total field. In fact, to distinguish it
from the latter, we call it the incident field and designate it by
$u^{i}$ such that
\begin{equation}\label{screen5}
u^{i}(\mathbf{x},\omega)=\int_{\Omega^{i}}G(\mathbf{x},\mathbf{x}',\omega)
s(\mathbf{x}',\omega)d\varpi(\mathbf{x}')
~~;~~\mathbf{x}\in \mathbb{R}^{2}~,
\end{equation}
wherein we have employed the fact that the support of the sources
is $\Omega^{i}\subset\mathbb{R}^{2}$.

The total field is then
$u(\mathbf{x},\omega)=u^{i}(\mathbf{x},\omega)+u^{d}(\mathbf{x},\omega)$
to the left of the screen, and
$u(\mathbf{x},\omega)=u^{d}(\mathbf{x},\omega)$ to the right of
the screen (other definitions of the diffracted field
$u^{d}(\mathbf{x},\omega)$ are, of course, possible, but we choose
this one). The diffracted field satisfies:
\begin{equation}\label{screen6}
(\triangle+k^{2})u^{d}(\mathbf{x},\omega)=0~~;~~\mathbf{x}\in
\Omega^{-}\cup\Omega^{+}\cup\Gamma^{0}~,
\end{equation}
\begin{equation}\label{screen7}
u^{d}(\mathbf{x},\omega)\sim \text{outgoing
waves}~;~\|\mathbf{x}\|\rightarrow\infty~~,~~\mathbf{x}\in\Omega^{-}~,~~\mathbf{x}\in~\Omega^{+}~.
\end{equation}
\begin{equation}\label{screen8}
u(c^{-},x_{2},\omega)=u(c^{+},x_{2},\omega)~~;~~x_{2}\in[-d/2,d/2]~,
\end{equation}
\begin{equation}\label{screen9}
u_{,2}(c^{-},x_{2},\omega)=u_{,2}(c^{+},x_{2},\omega)~~;~~x_{2}\in[-d/2,d/2]~,
\end{equation}
wherein $c^{\pm}:=\lim_{\epsilon\rightarrow 0}c\pm\epsilon$.  The
notion of impenetrability of the screen implies either
\begin{equation}\label{screen10}
u(c^{-},x_{2},\omega)=u(c^{+},x_{2},\omega)=0~~;~~x_{2}\in\mathbb{R}-[-d/2,d/2]~,
\end{equation}
(for a so-called acoustically-soft screen) or
\begin{equation}\label{screen11}
u_{,2}(c^{-},x_{2},\omega)=u_{,2}(c^{+},x_{2},\omega)=0~~;~~x_{2}\in\mathbb{R}-[-d/2,d/2]~,
\end{equation}
(for a so-called acoustically-hard screen). This notion of
impenetrability will be generalized further on.

We now apply Green's theorem to  the Green's function $G$ and
$u^{d}$ in $\Omega^{+}$ so as to obtain
\begin{equation}\label{screen12}
\mathcal{H}_{\Omega^{+}}(\mathbf{x})u^{d}(\mathbf{x},\omega)=\int_{\partial\Omega^{+}}\left[
G(\mathbf{x},\mathbf{x}',\omega)\boldsymbol{\nu}(\mathbf{x}')\cdot\nabla'
u^{d}(\mathbf{x}',\omega)-u^{d}(\mathbf{x}',\omega)\boldsymbol{\nu}(\mathbf{x}')\cdot\nabla'
G(\mathbf{x},\mathbf{x}',\omega)\right] d\gamma(\mathbf{x}')~.
\end{equation}
wherein $d\gamma(\mathbf{x}')$ is the infinitesimal element of arc
length,
$\partial\Omega^{+}:=\Gamma_{\infty}^{+}+\Gamma^{-}+\Gamma^{0}+\Gamma^{+}$,
and
\begin{equation}\label{screen13}
\mathcal{H}_{\Omega^{+}}(\mathbf{x})= \left \{
\begin{array}{cc}
1 & ~~;~\mathbf{x}\in\Omega^{+}
\\
0 &
~~~~~~~~~~~~~~;~\mathbf{x}\notin(\Omega^{+}+\partial\Omega^{+})
\end{array}\right.
~.
\end{equation}
On account of the radiation conditions satisfied by $G$ and
$u^{d}$, the integral along $\Gamma_{\infty}^{+}$ vanishes and
since $u^{d}=u$ on $\partial\Omega^{+}$ and in $\Omega^{+}$, we
have
\begin{equation}\label{screen14}
\mathcal{H}_{\Omega^{+}}(\mathbf{x})u(\mathbf{x},\omega)=\int_{\Gamma^{-}+\Gamma^{0}+\Gamma^{+}}\left[
G(\mathbf{x},\mathbf{x}',\omega)\boldsymbol{\nu}(\mathbf{x}')\cdot\nabla'
u(\mathbf{x}',\omega)-u(\mathbf{x}',\omega)\boldsymbol{\nu}(\mathbf{x}')\cdot\nabla'
G(\mathbf{x},\mathbf{x}',\omega)\right] d\gamma(\mathbf{x}')~,
\end{equation}
from which we extract the two results:
\begin{equation}\label{screen15}
u(\mathbf{x},\omega)=\int_{\Gamma^{-}+\Gamma^{0}+\Gamma^{+}}\left[
G(\mathbf{x},\mathbf{x}',\omega)\boldsymbol{\nu}(\mathbf{x}')\cdot\nabla'
u(\mathbf{x}',\omega)-u(\mathbf{x}',\omega)\boldsymbol{\nu}(\mathbf{x}')\cdot\nabla'
G(\mathbf{x},\mathbf{x}',\omega)\right]
d\gamma(\mathbf{x}')~;~\mathbf{x}\in\Omega^{+}~,
\end{equation}
\begin{equation}\label{screen16}
0=\int_{\Gamma^{-}+\Gamma^{0}+\Gamma^{+}}\left[
G(\mathbf{x},\mathbf{x}',\omega)\boldsymbol{\nu}(\mathbf{x}')\cdot\nabla'
u(\mathbf{x}',\omega)-u(\mathbf{x}',\omega)\boldsymbol{\nu}(\mathbf{x}')\cdot\nabla'
G(\mathbf{x},\mathbf{x}',\omega)\right]
d\gamma(\mathbf{x}')~;~\mathbf{x}\in\Omega^{-}~.
\end{equation}
Eq. (\ref{screen15}) is a bona fide boundary integral
representation of the field in the right hand half space. If the
integral in (\ref{screen16}) were equal to $u(\mathbf{x},\omega)$
in the left hand half space, then (\ref{screen16}) would seem to
imply that $u(\mathbf{x},\omega)=0~;~\mathbf{x}\in\Omega^{-}$
which would mean that, by some miracle, we had devised a
unilateral radiator by simply placing a screen with a slit in
front of an arbitrary source distribution. This, of course, cannot
be true, but it is a result that we are aiming for.

To proceed further in rigorous manner would require solving an
integral equation, a procedure we wish to avoid. Thus, we adopt
the approximation method employed since more than a hundred years
by many researchers in the acoustics, electromagnetics, and optics
communities (\nocite{baco50}Baker and Copson, 1950). To begin,
this involves the generalization of the notion of screen
impenetrability, which, simply stated, requires that {\it the
screen is simultaneously acoustically-hard and acoustically-soft}.
The consequence of this (mathematically-impossible) requirement is
that
\begin{equation}\label{screen17}
\mathcal{H}_{\Omega^{+}}(\mathbf{x})u(\mathbf{x},\omega)\approx\int_{\Gamma^{0}}\left[
G(\mathbf{x},\mathbf{x}',\omega)\boldsymbol{\nu}(\mathbf{x}')\cdot\nabla'
u(\mathbf{x}',\omega)-u(\mathbf{x}',\omega)\boldsymbol{\nu}(\mathbf{x}')\cdot\nabla'
G(\mathbf{x},\mathbf{x}',\omega)\right] d\gamma(\mathbf{x}')~,
\end{equation}
wherein we have replaced the previous $=$ sign by the $\approx$
sign to stress the fact that we are violating a mathematical
constraint (which is generally the case when some approximate
boundary conditions are invoked).

A further aspect of the procedure adopted by the above-mentioned
researchers is the introduction of the so-called {\it Kirchhoff
approximation} of the field in the slit. This ansatz (similar in
some respects to the Born approximation in other contexts) is
expressed by
\begin{equation}\label{screen18}
u(\mathbf{x},\omega)\approx
u^{i}(\mathbf{x},\omega)~~,~~\boldsymbol{\nu}(\mathbf{x})\cdot\nabla
u(\mathbf{x},\omega)\approx
\boldsymbol{\nu}(\mathbf{x})\cdot\nabla
u^{i}(\mathbf{x},\omega)~~;~~\mathbf{x}\in\Gamma^{0}~,
\end{equation}
which has been shown to be reasonable as soon as the width of the
slit exceeds several wavelengths (\nocite{faro72}Facq and Robin,
1972; \nocite{cofa73}Colombeau et al., 1973) and results in the
even-stronger approximation
\begin{equation}\label{screen19}
\mathcal{H}_{\Omega^{+}}(\mathbf{x})u(\mathbf{x},\omega)\approx\int_{\Gamma^{0}}\left[
G(\mathbf{x},\mathbf{x}',\omega)\boldsymbol{\nu}(\mathbf{x}')\cdot\nabla'
u^{i}(\mathbf{x}',\omega)-u^{i}(\mathbf{x}',\omega)\boldsymbol{\nu}(\mathbf{x}')\cdot\nabla'
G(\mathbf{x},\mathbf{x}',\omega)\right] d\gamma(\mathbf{x}')~,
\end{equation}
or, on account of the fact that
$\boldsymbol{\nu}(\mathbf{x}')\cdot\nabla'=-\frac{\partial}{\partial
x'_{1}}$ and $d\gamma(\mathbf{x}')=dx'_{2}$ along $\Gamma^{0}$:
\begin{equation}\label{screen20}
\mathcal{H}_{\Omega^{+}}(\mathbf{x})u(\mathbf{x},\omega)\approx-\int_{-d/2}^{d/2}\left[
G(\mathbf{x},c,x'_{2},\omega)
u_{,1'}^{i}(c,x'_{2},\omega)-u^{i}(c,x'_{2},\omega)
G_{,1'}(\mathbf{x},c,x'_{2},\omega)\right] dx'_{2}~.
\end{equation}
Once again, we extract two results from this expression:
\begin{equation}\label{screen21}
u(\mathbf{x},\omega)\approx-\int_{-d/2}^{d/2}\left[
G(\mathbf{x},c,x'_{2},\omega)
u_{,1'}^{i}(c,x'_{2},\omega)-u^{i}(c,x'_{2},\omega)
G_{,1'}(\mathbf{x},c,x'_{2},\omega)\right]
dx'_{2}~~;~~\mathbf{x}\in\Omega^{+}~,
\end{equation}
\begin{equation}\label{screen22}
0\approx-\int_{-d/2}^{d/2}\left[ G(\mathbf{x},c,x'_{2},\omega)
u_{,1'}^{i}(c,x'_{2},\omega)-u^{i}(c,x'_{2},\omega)
G_{,1'}(\mathbf{x},c,x'_{2},\omega)\right]
dx'_{2}~~;~~\mathbf{x}\in\Omega^{-}~.
\end{equation}
The previous remarks apply even more forcefully here.
%%%%%%%%%%%%%%%%%%%%%%%%%%%%%%%%%%%%%%
\section{A radiation problem with a particular type and distribution of
sources}\label{synth}
The previous analysis showed that  employing the generalized
impenetrability conditions and the Kirchoff approximation enables
one to obtain an approximate solution for the field in the right
hand half space and what appears like a null field in the left
hand half space around a slit in an impenetrable screen. Since
this result is approximate, it does not satisfy the governing
equations of the original problem. As concerns the space-frequency
wave equation expressed in (\ref{screen6}), this fact is easy to
demonstrate by simply taking the spatial derivatives of
(\ref{screen20}) whereupon one finds that the right hand side of
the equation in (\ref{screen6}) is no longer nil, i.e.,
\begin{equation}\label{synth1}
(\triangle+k^{2})u(\mathbf{x},\omega)=-S(\mathbf{x},\omega)\neq
0~~;~~\mathbf{x}\in\mathbb{R}^{2}~.
\end{equation}
Rather than do these operations, we will give the result for $S$
and then go the other way around by showing that $u$ for this
source distribution has the desired properties, notably of
producing a null field in $\Omega^{-}$.

The source density ansatz is
\begin{multline}\label{synth2}
S(\mathbf{x},\omega)=-2\delta(x_{1}-c)[H(x_{2}-d/2)+H(-x_{2}-d/2)]u_{,1}^{i}(\mathbf{x},\omega)-
\\
u^{i}(\mathbf{x},\omega)\delta_{,1}(x_{1}-c)[H(x_{2}-d/2)+H(-x_{2}-d/2)]
~~;~~\mathbf{x}\in\mathbb{R}^{2}~,
\end{multline}
(wherein $H$ is the Heaviside function defined by $H(\zeta>0)=1$
and $H\zeta<0)=0$) which will be recognized to be a distribution
of single and double sources on a strip. The latter is none other
than the slit of the previous problem.

To ensure uniqueness of this radiation problem, we must specify
that the wave radiated by this distribution of sources is outgoing
far from the support of the sources.

As previously, we can show that the radiated field is of the form
\begin{equation}\label{synth3}
u(\mathbf{x},\omega)=\int_{\mathbb{R}^{2}}G(\mathbf{x},\mathbf{x}',\omega)
S(\mathbf{x}',\omega)d\varpi(\mathbf{x}')
~~;~~\mathbf{x}\in\mathbb{R}^{2}~,
\end{equation}
or
\begin{equation}\label{synth4}
u(\mathbf{x},\omega)=\int_{-\infty}^{\infty}dx'_{2}\int_{-\infty}^{\infty}dx'_{1}
G(\mathbf{x},x'_{1},x'_{2},\omega)S(x'_{1},x'_{2},\omega)
~~;~~\mathbf{x}\in\mathbb{R}^{2}~.
\end{equation}
The introduction of (\ref{synth2}) therein gives
\begin{multline}\label{synth5}
u(\mathbf{x},\omega)=\int_{-\infty}^{\infty}dx'_{2}[H(x'_{2}-d/2)+H(-x'_{2}-d/2)]
\times
\\
 \int_{-\infty}^{\infty}dx'_{1}
G(\mathbf{x},x'_{1},x'_{2},\omega)[
-2\delta(x'_{1}-c)u_{,1'}^{i}(\mathbf{x}',\omega)-u^{i}(\mathbf{x}',\omega)\delta_{,1'}(x'_{1}-c)]
~~;~~\mathbf{x}\in\mathbb{R}^{2}~,
\end{multline}
or
\begin{equation}\label{synth6}
u(\mathbf{x},\omega)=\int_{-d/2}^{d/2}dx'_{2}
 \int_{-\infty}^{\infty}dx'_{1}
G(\mathbf{x},x'_{1},x'_{2},\omega)[
-2\delta(x'_{1}-c)u_{,1'}^{i}(\mathbf{x}',\omega)-u^{i}(\mathbf{x}',\omega)\delta_{,1'}(x'_{1}-c)]
~~;~~\mathbf{x}\in\mathbb{R}^{2}~.
\end{equation}
But
\begin{equation}\label{synth7}
 \int_{-\infty}^{\infty}dx'_{1}
G(\mathbf{x},x'_{1},x'_{2},\omega)\delta(x'_{1}-c)u_{,1'}^{i}(\mathbf{x}',\omega)=
G(\mathbf{x},c,x'_{2},\omega)u_{,1'}^{i}(c,x'_{2},\omega)
~~;~~\mathbf{x}\in\mathbb{R}^{2}~.
\end{equation}
Furthermore:
\begin{multline}\label{synth8}
 \int_{-\infty}^{\infty}dx'_{1}
G(\mathbf{x},x'_{1},x'_{2},\omega)u^{i}(\mathbf{x}',\omega)\delta_{,1'}(x'_{1}-c)=
G(\mathbf{x},x'_{1},x'_{2},\omega)\delta(x'_{1}-c)u^{i}(x'_{1},x'_{2},\omega)\Big
| _{x'_{1}= -\infty}^{\infty}-
\\
\int_{-\infty}^{\infty}[G(\mathbf{x},x'_{1},x'_{2},\omega)u^{i}(\mathbf{x}',\omega)]_{,1'}
\delta(x'_{1}-c)dx'_{1}~~;~~\mathbf{x}\in\mathbb{R}^{2}~.
\end{multline}
But $\delta(\pm\infty-c)=0$ and
$G(\mathbf{x},\pm\infty,x'_{2},\omega)u^{i}(\pm\infty,x'_{2},\omega)$
is bounded, so that
\begin{multline}\label{synth9}
 \int_{-\infty}^{\infty}dx'_{1}
G(\mathbf{x},x'_{1},x'_{2},\omega)u^{i}(\mathbf{x}',\omega)\delta_{,1'}(x'_{1}-c)=
\\
-\int_{-\infty}^{\infty}[G(\mathbf{x},x'_{1},x'_{2},\omega)u_{,1'}^{i}(\mathbf{x}',\omega)+
G_{,1'}(\mathbf{x},x'_{1},x'_{2},\omega)u^{i}(\mathbf{x}',\omega)]
\delta(x'_{1}-c)dx'_{1}=
\\
-G(\mathbf{x},c,x'_{2},\omega)u_{,1'}^{i}(c,x'_{2},\omega)-
G_{,1'}(\mathbf{x},c,x'_{2},\omega)u^{i}(c,x'_{2},\omega)
~~;~~\mathbf{x}\in\mathbb{R}^{2}~.
\end{multline}
The introduction of (\ref{synth7}) and (\ref{synth9}) into
(\ref{synth6}) then gives
\begin{multline}\label{synth10}
u(\mathbf{x},\omega)=\int_{-d/2}^{d/2}dx'_{2}
 \Big[
 -2G(\mathbf{x},c,x'_{2},\omega)u_{,1'}^{i}(c,x'_{2},\omega)+
 \\
 G(\mathbf{x},c,x'_{2},\omega)u_{,1'}^{i}(c,x'_{2},\omega)+
G_{,1'}(\mathbf{x},c,x'_{2},\omega)u^{i}(c,x'_{2},\omega)\Big]
~~;~~\mathbf{x}\in\mathbb{R}^{2}~,
\end{multline}
or
\begin{equation}\label{synth11}
u(\mathbf{x},\omega)=-\int_{-d/2}^{d/2}
 \Big[
 G(\mathbf{x},c,x'_{2},\omega)u_{,1'}^{i}(c,x'_{2},\omega)-
G_{,1'}(\mathbf{x},c,x'_{2},\omega)u^{i}(c,x'_{2},\omega)\Big]
dx'_{2} ~~;~~\mathbf{x}\in\mathbb{R}^{2}~,
\end{equation}
from which we extract the two ({\it rigorous}) results:
\begin{equation}\label{synth12}
u(\mathbf{x},\omega)=-\int_{-d/2}^{d/2}
 \Big[
 G(\mathbf{x},c,x'_{2},\omega)u_{,1'}^{i}(c,x'_{2},\omega)-
G_{,1'}(\mathbf{x},c,x'_{2},\omega)u^{i}(c,x'_{2},\omega)\Big]
dx'_{2} ~~;~~\mathbf{x}\in\Omega^{+}~,
\end{equation}
\begin{equation}\label{synth13}
u(\mathbf{x},\omega)=-\int_{-d/2}^{d/2}
 \Big[
 G(\mathbf{x},c,x'_{2},\omega)u_{,1'}^{i}(c,x'_{2},\omega)-
G_{,1'}(\mathbf{x},c,x'_{2},\omega)u^{i}(c,x'_{2},\omega)\Big]
dx'_{2} ~~;~~\mathbf{x}\in\Omega^{-}~.
\end{equation}
We showed previously (see (\ref{screen22})) that
\begin{equation}\label{synth14}
-\int_{-d/2}^{d/2}\left[ G(\mathbf{x},c,x'_{2},\omega)
u_{,1'}^{i}(c,x'_{2},\omega)-u^{i}(c,x'_{2},\omega)
G_{,1'}(\mathbf{x},c,x'_{2},\omega)\right] dx'_{2}\approx 0
~~;~~\mathbf{x}\in\Omega^{-}~.
\end{equation}
so that we can conclude that
\begin{equation}\label{synth15}
u(\mathbf{x},\omega)\approx 0 ~~;~~\mathbf{x}\in\Omega^{-}~,
\end{equation}
where it is understood that the field $u(\mathbf{x},\omega)$ in
(\ref{synth12}) and (\ref{synth15}) is the field radiated by the
distribution of applied sources given in (\ref{synth2}). {\it This
means that the source function (\ref{synth2}) radiates in the
sought-for unilateral manner}. Of course, this is only an
approximation, but we shall discover further on that it is a good
approximation. Thus, we have shown how to synthesize the sources
that give rise to unilateral radiator. Replacing the physical
radiator (e.g., parabolic antenna, cylindrical transducer) by this
source distribution enables the elimination of undesirable
multiple diffraction effects between the radiator and obstacles.
%%%%%%%%%%%%%%%%%%%%%%%%%%%%%%%%%%%%%%%%%%%%%%%%%%%%%%%%%%%%%%%%%
\section{Radiation from a parabolic cylinder
radiator}\label{parabol}
The cross section view of the parabolic cylinder radiator is given
in fig. \ref{fig0}. An infinitely-thin impenetrable curved sheet
($\Gamma$) reflector, in the form of a portion of a parabola, is
illuminated by a line source located at the origin $O$. The
equation of the sheet is
\begin{equation}\label{parabol1}
x_{1}=F(x_{2}):=-f+\frac{x_{2}^{2}}{4f}~~;~~x_{2}\in[-d/2,d/2]~,
\end{equation}
wherein $f$ is the focal length and $d$ the width of the reflector
aperture. The left hand extremity of the reflector is at
$x_{1}=-f$ and the right hand extremity at $x_{1}=c$. Thus, the
slit aperture $\mathcal{A}$ is located at
$x_{1}=c=-f+\frac{d^{2}}{16f}$.

The problem is once again to determine the total field
$U=U^{i}+U^{d}$ such that
\begin{equation}\label{parabol1a}
(\triangle+k^{2})U^{i}(\mathbf{x},\omega)=-\delta(\mathbf{x})~~;~~\mathbf{x}\in\mathbb{R}^{2}~,
\end{equation}
\begin{equation}\label{parabol2}
(\triangle+k^{2})U^{d}(\mathbf{x},\omega)=0~~;~~\mathbf{x}\in\mathbb{R}^{2}\cap\Gamma~,
\end{equation}
\begin{equation}\label{parabol3}
U^{d}(\mathbf{x},\omega)\sim \text{outgoing
waves}~;~\|\mathbf{x}\|\rightarrow\infty~~,~~\mathbf{x}\in\mathbb{R}^{2}~.
\end{equation}
\begin{equation}\label{parabol4}
U(\mathbf{x},\omega)=0~~;~~\mathbf{x}\in\Gamma~.
\end{equation}
Note that the Dirichlet boundary condition (\ref{parabol4}) could
just as well be replaced by the Neumann condition
\begin{equation}\label{parabol5}
\boldsymbol{\nu}\cdot \nabla
U(\mathbf{x},\omega)=0~~;~~\mathbf{x}\in\Gamma~,
\end{equation}
or, for that matter, by an impedance boundary condition, since, in
the high frequency situation of interest herein, the precise
nature of the boundary condition is not important.

There exists a variety of exact and approximate methods for
predicting the radiation produced by this device
(\nocite{tawi75}Tanteri and Wirgin, 1975); we shall choose a
so-called {\it aperture method} (\nocite{si49}Silver, 1949) which
is approximate in nature and based on the following two
hypotheses:
\begin{enumerate}
\item geometrical optics (or acoustics) governs the propagation of
the field from the source to the reflector and from the latter to
the aperture,
\item the Huyghens-Fresnel principle (equivalent to the result of the analysis in sect. \ref{screen})
governs the propagation of the
field from the aperture of the mirror to points within
$\Omega^{+}$ (i.e., the half space to the right of the aperture).
\end{enumerate}
A necessary (although not necessarily-sufficient) condition for
the validity of the first hypothesis is that $kf>>1$
(\nocite{tawi75}Tanteri and Wirgin, 1975). If it is recalled that
the field radiated by the line source located at $\mathbf{x}'=0$
is $\frac{i}{4}H_{0}^{(1)}(k\|\mathbf{x}\|)$ and that the
asymptotic (large-argument) form of the Hankel function is
(\nocite{abst65}Abramowitz and Stegun, 1965)
\begin{equation}\label{parabol6}
H_{n}^{(1)}(\zeta)\sim\left(
\frac{2}{\pi\zeta}\right)^{\frac{1}{2}}e^{i\left(
\zeta-\frac{n\pi}{2}-\frac{\pi}{4}\right) }
~~;~~\zeta\rightarrow\infty~~,~~n=0,1,2,...~,
\end{equation}
then the incident field at the reflector can be replaced by the
asymptotic expression
\begin{equation}\label{parabol7}
U^{i}(\mathbf{x},\omega)\sim \tilde{U}^{i}(\mathbf{x},\omega)=
\frac{i\xi}{4}\frac{e^{ikfE(x_{2})}}{\left[kfE(x_{2})\right]
^{\frac{1}{2}} }~~;~~\mathbf{x}\in\Gamma~,
\end{equation}
wherein $\xi:=\sqrt{\frac{2}{\pi}}e^{-i\frac{\pi}{4}}$ and
$E(x_{2}):=1+\frac{x_{2}^{2}}{4f^{2}}$. The standard geometrical
optics (acoustics) ray analysis (\nocite{sl69}Sletten, 1969;
\nocite{ho69}Holt, 1969; \nocite{ha64}Hansen, 1964;
\nocite{si49}Silver, 1949) then shows that
\begin{equation}\label{parabol8}
U^{d}(\mathbf{x},\omega)\sim
\tilde{U}^{d}(\mathbf{x},\omega)=\left \{
\begin{array}{cc}
A(x_{2})\exp[i(kx_{1}+\psi)] & ~;~\mathbf{x}\in\tilde{\Omega}
\\
0 & ~~~~~~~~~~~~~~~;~\mathbf{x}\in
(\mathbb{R}^{2}-\Gamma-\tilde{\Omega})
\end{array}\right.
~,
\end{equation}
wherein
\begin{equation}\label{parabol9}
\tilde{\Omega}=\{F(x_{2})<x_{1}<\infty~~;~~\forall x_{2}\in
[-d/2,d/2]\}~,
\end{equation}
\begin{equation}\label{parabol10}
A(x_{2})=\|\tilde{U}^{i}(F(x_{2}),x_{2},\omega)\|=\frac{1}{4}\sqrt{\frac{2}{\pi
kf}}\frac{1}{\sqrt{E(x_{2})}}~,
\end{equation}
\begin{equation}\label{parabol11}
\psi=\arg[-\tilde{U}^{i}(F(x_{2}),x_{2},\omega)]+kf=-\frac{3\pi}{4}+2kf
~.
\end{equation}
Thus, the first hypothesis of this aperture method, is equivalent
to the statement that the phase $kc+\psi$ is a constant, and the
amplitude $A(x_{2})$ is a tapered function of $x_{2}$, in the
aperture $\mathcal{A}$ of the parabolic reflector. Note that for a
plane body wave normally-incident on the screen+slit system, both
the amplitude and  phase are constant in $\mathcal{A}$.

Naturally, (\ref{parabol8}) does not account for diffraction
effects (assumed to be produced only in the half space to the
right of the aperture) due to the encounter of $u^{i}$ with the
edges of the mirror. The contribution of these effects to points
within $\Omega^{+}$ is introduced by means of the second
hypothesis whose mathematical expression is given either by the
first or second Rayleigh-Sommerfeld formulae (\nocite{si49}Siver,
1949), or, as herein, by (\ref{screen21}). The choice of one or
another of these formulae constitutes the essential difference
between the three types of aperture methods. Experience shows that
the two Rayleigh-Sommerfled formulae yield similar results in both
the near and far field regions of $\Omega^{+}$ when $kf>2.5\pi$.
We prefer the third aperture method, embodied in (\ref{screen21}),
because it has the unique property of leading to unilateral
radiation if abstraction is made of the reflector antenna once the
field it generates attains the aperture.

The question that arises is what should be taken for $u^{i}$
and/or $u_{,1'}^{i}$ in (\ref{screen21}). In other words, should
one take $u^{i}\approx\tilde{U}$ and
$u^{i}_{,1'}\approx\tilde{U}_{,1'}$, or, on the contrary,
$u^{i}\approx\tilde{U}^{d}$ and
$u^{i}_{,1'}\approx\tilde{U}^{d}_{,1'}$? Due to the fact that, in
practice, the source of reflector antennas is generally masked so
as not to radiate in directions other than towards the reflector,
it seems reasonable to choose the second solution.

Thus, we take
\begin{equation}\label{parabol12}
u(\mathbf{x},\omega)\approx-\int_{-d/2}^{d/2}\left[
G(\mathbf{x},c,x'_{2},\omega)
\tilde{U}_{,1'}^{d}(c,x'_{2},\omega)-\tilde{U}^{d}(c,x'_{2},\omega)
G_{,1'}(\mathbf{x},c,x'_{2},\omega)\right]
dx'_{2}~~;~~\mathbf{x}\in\Omega^{+}~,
\end{equation}
wherein
\begin{equation}\label{parabol13}
\tilde{U}^{d}(c,x'_{2},\omega)=A(x'_{2})\exp[i(kc+\psi)]~~;~~x_{2}\in[-d/2,d/2]~,
\end{equation}
\begin{equation}\label{parabol13a}
\tilde{U}_{,1'}^{d}(c,x'_{2},\omega)=ikA(x'_{2})\exp[i(kc+\psi)]~~;~~x_{2}\in[-d/2,d/2]~.
\end{equation}
Although we are now in a position to compute the field radiated
into $\Omega^{+}$, we shall not accomplish this task since what we
are really interested in is the prediction of the radiation from
the aperture sources that synthesize the action of a parabolic
cylinder unilateral radiator.
%%%%%%%%%%%%%%%%%%%%%%%%%%%%%%%%%%%%%
\section{The quasi-unilateral radiation from the synthesized sources of a parabolic
cylinder antenna}\label{anten}
By taking into the final results of sects. \ref{synth} and
\ref{parabol}, we find that the field radiated by the sources
which synthesize the action of a parabolic cylinder antenna is
\begin{equation}\label{anten1}
u(\mathbf{x},\omega)\approx-\int_{-d/2}^{d/2}\left[
G(\mathbf{x},c,x'_{2},\omega)
\tilde{U}_{,1'}^{d}(c,x'_{2},\omega)-\tilde{U}^{d}(c,x'_{2},\omega)
G_{,1'}(\mathbf{x},c,x'_{2},\omega)\right]
dx'_{2}~~;~~\mathbf{x}\in\mathbb{R}^{2}~,
\end{equation}
wherein $\tilde{U}^{d}$ and $\tilde{U}_{,1'}^{d}$ are given in
(\ref{parabol12})-(\ref{parabol13}) and
\begin{equation}\label{anten2}
G(\mathbf{x},\mathbf{x}',\omega)=\frac{i}{4}H_{0}^{(1)}
(k|\sqrt{(x_{1}-x'_{1})^{2}+(x_{2}-x'_{2})^{2}}|)~,
\end{equation}
so that
\begin{equation}\label{anten3}
G(\mathbf{x},c,x'_{2},\omega)=\frac{i}{4}H_{0}^{(1)}(k|R|)~,
\end{equation}
wherein  $R=\sqrt{(x_{1}-c)^{2}+(x_{2}-x'_{2})^{2}}$, and
\begin{equation}\label{anten4}
G_{,1'}(\mathbf{x},c,x'_{2},\omega)=
\frac{ik}{4}\frac{(x_{1}-c)}{|R|}H_{1}^{(1)}(k|R|)~.
\end{equation}
It follows that
\begin{equation}\label{anten5}
u(\mathbf{x},\omega)\approx \frac{k}{4}
e^{i(kc+\psi)}\int_{-d/2}^{d/2}A(x'_{2})\left[
H_{0}^{(1)}(k|R|)+i\frac{(x_{1}-c)}{|R|}H_{1}^{(1)}(k|R|) \right]
dx'_{2}~~;~~\mathbf{x}\in\mathbb{R}^{2}~.
\end{equation}
This expression can be computed by any (e.g., rectangle) numerical
quadrature scheme.
%%%%%%%%%%%%%%%%%%%%%%%%
\subsection{Far field zone radiation}
Let  $r',~\phi'$ be the polar coordinates of the integration point
subtended by the vector $\mathbf{x'}$ such that $r'\cos\phi'=c$,
and  $r,~\phi$ the polar coordinates of the observation point
subtended by the vector $\mathbf{x}$ such that $r\cos\phi=x_{1}$.
Then
\begin{equation}\label{anten6}
R=\sqrt{r^{2}+r'^{2}-2rr'\cos(\phi-\phi')}~.
\end{equation}
In the far field (Fraunhofer) zone, $r>>1$, and we assume also
that $kr>>1$, from which it follows that $k|R|>>1~;~\forall
x_{2}'\in[-d/2,d/2]$, so that we can make use of the asymptotic
forms
\begin{equation}\label{anten7}
H_{0}^{(1)}(k|R|)\sim\left( \frac{2}{\pi
k|R|}\right)^{\frac{1}{2}}e^{i\left( k|R|-\frac{\pi}{4}\right)
}~~,~~H_{1}^{(1)}(k|R|)\sim\left( \frac{2}{\pi
k|R|}\right)^{\frac{1}{2}}e^{i\left(
k|R|-\frac{\pi}{2}-\frac{\pi}{4}\right) }
~~;~~k|R|\rightarrow\infty~,
\end{equation}
so as to obtain
\begin{equation}\label{anten8}
u(\phi)\approx\frac{k}{4}
e^{i(kc+\psi)}\int_{-d/2}^{d/2}A(x'_{2})\sqrt{\frac{2}{\pi
k|R|}}e^{i\left( k|R|-\frac{\pi}{4}\right)}\left[ 1+
\frac{(x_{1}-c)}{|R|} \right]
dx'_{2}~~;~~\mathbf{x}\in\mathbb{R}^{2}~.
\end{equation}
Furthermore:
\begin{equation}\label{anten9}
R\approx r-r'\cos(\phi-\phi')~~;~~\frac{r'}{r}<<1~,
\end{equation}
so that we can make the approximations
\begin{equation}\label{anten9a}
\frac{1}{\sqrt{|R|}}\approx
\frac{1}{\sqrt{r}}~~,~~\frac{x_{1}-c}{|R|\sqrt{|R|}}\approx\cos\phi~~,~~e^{i
k|R|}\approx e^{ik[r-r'\cos(\phi-\phi')]} ~,
\end{equation}
 Then
\begin{equation}\label{anten10}
u(\mathbf{x},\omega)\sim \hat{u}(\phi)\left( \frac{2}{\pi
kr}\right) ^{\frac{1}{2}} e^{i\left( kr-\frac{\pi}{4}\right) }
~~;~~kr\rightarrow\infty~.
\end{equation}
and $\hat{u}(\phi)$ is the so-called {\it far field radiation
pattern} given by
\begin{equation}\label{anten11}
 \hat{u}(\phi)\approx\frac{k}{4}(1+\cos\phi)
e^{i[kc(1-\cos\phi)+\psi]}\int_{-d/2}^{d/2}A(x'_{2})e^{-ik
x'_{2}\sin\phi} dx'_{2}~~;~~\phi\in[0,2\pi[~.
\end{equation}
We note that in observation angles close to $\phi=\pi$ (the
backraditon angle), $\cos\phi\approx -1$, so that
$(1+\cos\phi)\approx 0$, which is the reason why this synthesized
source gives rise (approximately) to unilateral radiation.

\begin{figure}
[ptb]
\begin{center}
\includegraphics[scale=0.8] {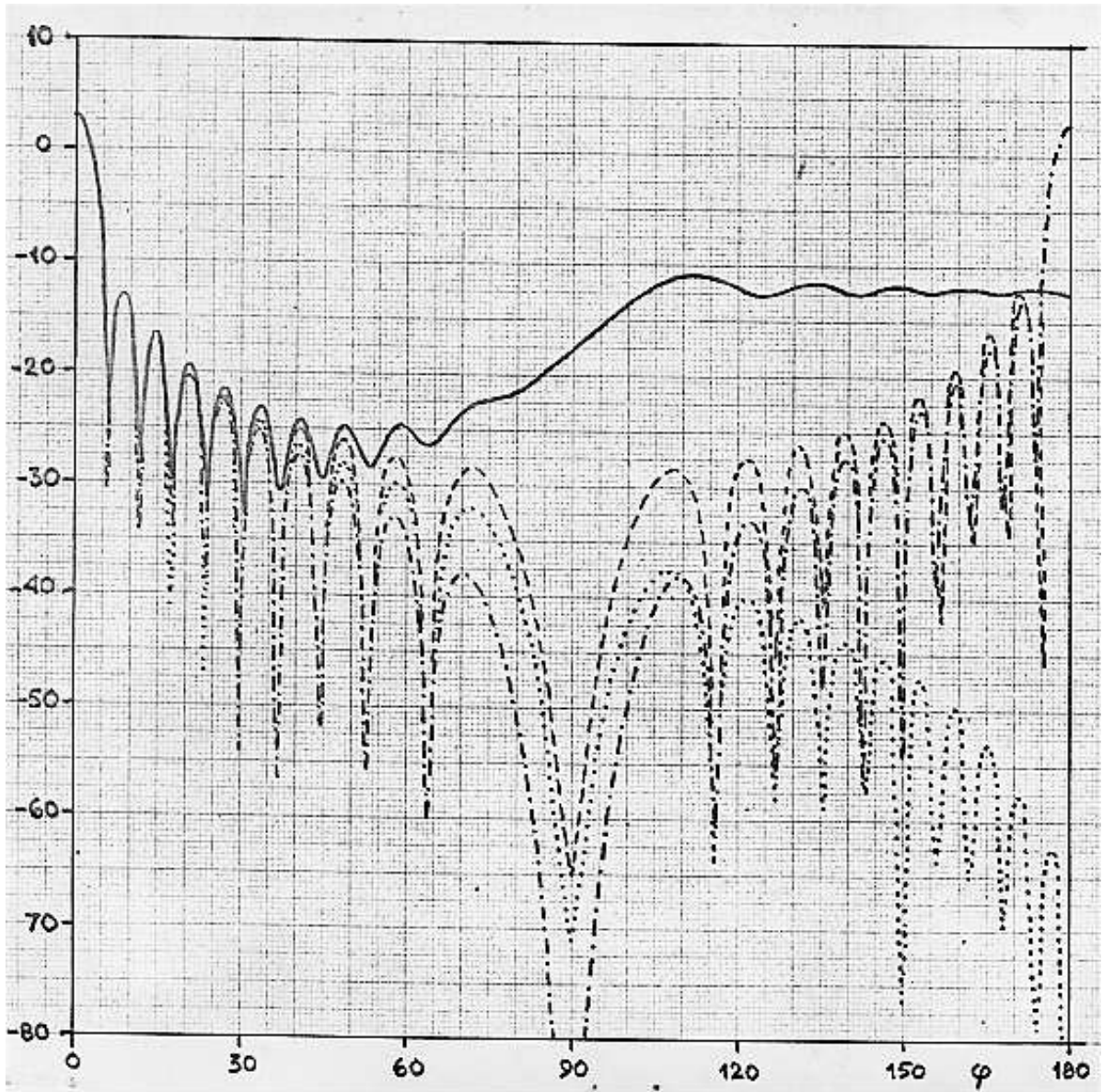}
  \caption{Graphs of the far-field  radiated power pattern $\sigma(\phi)$ for
  a parabolic cylinder radiator. The full line curve stems from the exact theory.
  The other three curves stem from the various aperture method descriptions of the
  action of the antenna. The dotted curve results from the aperture method relying
  on the  synthesized sources for unilateral radiation.}
  \label{fig3}
  \end{center}
\end{figure}

We can define the power radiation pattern in the Fraunhofer zone
by
\begin{equation}\label{anten12}
 \sigma(\phi):=10\log_{10}(\|\hat{u}(\phi)\|^{2}) \text{(in~dB)}~.
\end{equation}
We plot this function (dotted curve) in fig. \ref{fig3} for a
source distribution which synthesizes the radiation of a parabolic
cylinder antenna for which $kd=20\pi$ and $c=0$. The full line
curve therein is the result of a rigorous computation of the field
radiated by this antenna on which a Dirichlet boundary condition
is imposed, and the other two curves represent the predictions
resulting from the other two aperture methods. We note that our
synthesized sources indeed give rise to very weak radiation in the
left hand half space (i.e., for $\phi>90^{o}$, there being
symmetry around $\phi=0$. We also not that the other three
radiation patterns do not possess this property.
%%%%%%%%%%%%%%%%%%%%%%%
%%
%%
%%%%%%%%%%%%%%%%%%%%%%%%%%%%%%%%%%%%%%%%%%%%%%
\section{Final comments on the use of the synthesized sources of a unilateral radiator in a
scattering problem}\label{scat}
A typical problem of the scattering of a  wave $u^{i}$ radiated
from synthesized sources $s$ of support $\Omega^{i}$ by some
boundary $\Gamma$ on which the field is nil (i.e., Dirichlet
boundary condition) is expressed as follows: determine the total
field $u=u^{i}+u^{d}$ in a domain $\Omega$ such that:
\begin{equation}\label{scat1}
(\triangle+k^{2})u(\mathbf{x},\omega)=-s(\mathbf{x})~~;~~\mathbf{x}\in\Omega~,
\end{equation}
\begin{equation}\label{scat2}
u^{d}(\mathbf{x},\omega)\sim \text{outgoing
waves}~;~\|\mathbf{x}\|\rightarrow\infty~~,~~\mathbf{x}\in\Omega~,
\end{equation}
\begin{equation}\label{scat3}
u(\mathbf{x},\omega)=0~~;~~\mathbf{x}\in\Gamma~,
\end{equation}
wherein
\begin{multline}\label{scat4}
s(\mathbf{x},\omega)=-2\delta(x_{1}-c)[H(x_{2}-d/2)+H(-x_{2}-d/2)]\tilde{U}_{,1}^{i}(\mathbf{x},\omega)-
\\
\tilde{U}^{i}(\mathbf{x},\omega)\delta_{,1}(x_{1}-c)[H(x_{2}-d/2)+H(-x_{2}-d/2)]
~~;~~\mathbf{x}\in\mathbb{R}^{2}~.
\end{multline}
and
\begin{equation}\label{scat5}
\tilde{U}^{i}(\mathbf{x},\omega)=A(x_{2})e^{i(kx_{1}+\psi)}~.
\end{equation}
Actually, it is possible to assume other expressions for
$\tilde{U}^{i}$ as long as they are connected in some plausible
way with an actual physically-realizable unilateral radiator.

The previous analysis showed that
\begin{equation}\label{scat6}
u^{i}(\mathbf{x},\omega)=\int_{\Omega^{i}}G(\mathbf{x},\mathbf{x}',\omega)
s(\mathbf{x}',\omega)d\varpi(\mathbf{x}')
~~;~~\mathbf{x}\in\Omega~,
\end{equation}
or, on account of (\ref{scat3}) (and whatever the expression for
$\tilde{U}^{i}$),
\begin{equation}\label{scat7}
u^{i}(\mathbf{x},\omega)=-\int_{-d/2}^{d/2}
 \Big[
 G(\mathbf{x},c,x'_{2},\omega)\tilde{U}_{,1'}^{i}(c,x'_{2},\omega)-
G_{,1'}(\mathbf{x},c,x'_{2},\omega)\tilde{U}^{i}(c,x'_{2},\omega)\Big]
dx'_{2} ~~;~~\mathbf{x}\in\Omega^{-}~.
\end{equation}
So much for the incident field on the boundary.

The next step is to find an appropriate (boundary integral, domain
integral, partial wave, etc.) representation of the scattered
field $u^{d}$ that incorporates the radiation condition
(\ref{scat2}). This involves some unknown functions that are
determined in the final step by application of the boundary
condition (\ref{scat3}).

It will be noted that the use of our synthesized source
distribution $s$: 1) enables us to simulate a unilateral radiated
incident field, and 2) obviates multiple diffraction between the
material boundaries of the radiator (of which abstraction is made
in this method) and those of the scatterer.
%%%%%%%%%%%%%%%%%%%%%%%%%%%%%%%%%%%%%%%%%%%%%%%%%%%%%%%%%%%%%%%%%%%%%%%
{\small\bibliographystyle{unsrt}
\bibliography{biblio_screen}}
%%%%%%%%%%%%%%%%%%%%%%%%%%%%%%%%%%%%%%%%%%%%%%%%%%%%%%%%%%%%%%%%%%%%%%%%%
\end{document}